# Superconductivity in single-crystalline, aluminum- and gallium-hyperdoped germanium

Slawomir Prucnal[1,*], Viton Heera[1], René Hübner[1], Mao Wang[1], Grzegorz P. Mazur[2,3], Michał J. Grzybowski[2,3], Xin Qin[4], Ye Yuan[1,5], Matthias Voelskow[1], Wolfgang Skorupa[1], Lars Rebohle[1], Manfred Helm[1], Maciej Sawicki[3], Shengqiang Zhou[1]

*1. Helmholtz-Zentrum Dresden-Rossendorf, Institute of Ion Beam Physics and Materials Research, Bautzner Landstraße 400, D-01328 Dresden, Germany*
*2. Institute of Physics, Polish Academy of Sciences, Aleja Lotnikow 32/46, PL-02668 Warsaw, Poland*
*3. International Research Centre MagTop, Institute of Physics, Polish Academy of Sciences, Aleja Lotnikow 32/46, PL-02668 Warsaw, Poland*
*4. Hefei National Laboratory for Physical Sciences at the Microscale, University of Science and Technology of China, Hefei, Anhui, 230026, China.*
*5. Physical Science and Engineering Division, King Abdullah University of Science and Technology, 23955-6900 Thuwal, Saudi Arabia*
*[*] corresponding author: s.prucnal@hzdr.de*

Abstract

Superconductivity in group IV semiconductors is desired for hybrid devices combining both semiconducting and superconducting properties. Following boron doped diamond and Si, superconductivity has been observed in gallium doped Ge, however the obtained specimen is in polycrystalline form [Herrmannsdörfer et al., Phys. Rev. Lett. 102, 217003 (2009)]. Here, we present superconducting single-crystalline Ge hyperdoped with gallium or aluminium by ion implantation and rear-side flash lamp annealing. The maximum concentration of Al and Ga incorporated into substitutional positions in Ge is eight times higher than the equilibrium solid solubility. This corresponds to a hole concentration above $10^{21}$ cm$^{-3}$. Using density functional theory in the local density approximation and pseudopotential plane-wave approach, we show that the superconductivity in p-type Ge is phonon-mediated. According to the *ab initio* calculations the critical superconducting temperature for Al- and Ga-doped Ge is in the range of 0.45 K for 6.25 at.% of dopant concentration being in a qualitative agreement with experimentally obtained values.

## 1. Introduction

Since the discovery of superconductivity in diamond [1] with a boron content above the equilibrium solid solubility (ESS) many studies have been performed to find new "superconducting semiconductors". Such a materials class would enable the monolithic integration of quantum and conventional electronics [2]. Indeed, several groups found superconductivity, even in the technologically more relevant semiconductors like Si [3], Ge [4] or SiC [5], after a heavy hole doping. A brief introduction into the research field of superconducting semiconductors was given in recent review articles [6-9].

The term "superconducting semiconductor" is a bit misleading since in a semiconductor, the carriers necessary for the Cooper pair condensate freeze out at low temperatures and superconductivity is impossible. Therefore, the semiconductor has to be heavily doped above the metal-insulator transition (MIT). It turned out that an acceptor concentration in excess of 1 at. % (i.e. above $5 \times 10^{20}$ cm$^{-3}$) is required to induce the superconductivity in germanium. Such concentration is higher than the ESS of typical acceptors in Ge. Hyperdoping, however, is difficult to achieve and requires non-equilibrium doping techniques, like a high-pressure high-temperature synthesis [1] or the chemical vapor deposition [8] in the case of diamond, gas immersion laser doping [3, 10, 11] and high-fluence ion implantation combined with rapid thermal annealing (RTA) or flash lamp annealing (FLA) [4, 12-15] for Si and Ge. Among these doping schemes ion implantation followed by FLA is best adopted to the current semiconductor technology.

Despite advanced non-equilibrium doping techniques, hyperdoped semiconductors are in most cases inhomogeneous materials with dopant concentration fluctuations [16] up to cluster or nanoprecipitate formation [13]. Moreover, dopant segregation at grain boundaries in polycrystalline materials or at interfaces to technologically relevant capping layers is a next serious problem [17]. There is an experimental evidence that in some semiconductor-acceptor systems, like Si:Ga, amorphous acceptor-rich nanoprecipitates ($c_{Ga}$ > 20 at.%) are vital for superconductivity [13]. Granularities of the superconducting condensates have been also obtained in boron doped diamond [18, 19]. Such granular superconductors can be modelled by a random network of Josephson junctions and exhibit a superconductor-insulator transition [20], as observed e.g. in Si:Ga [21, 22]. The presence of the superconductor-insulator transition clearly reveals the inhomogeneous character of the superconductor. Due to local superconducting regions, even in the insulating state such hyperdoped semiconductors demonstrate nonlinear transport phenomena [22] and anomalous large magnetoresistance [23].

However, for a perfect monolithic integration of superconducting nanocircuits in semiconductor devices, a homogeneous and single-crystalline structure is desirable. It remains an unresolved question whether superconducting semiconductor films of sufficient quality can be fabricated at all by today's top-down selective doping technologies and which semiconductor-acceptor combination is most promising. Since the tendency for disorder and cluster formation by hyperdoping increases with the covalent bond strength of the semiconductor and decreases with growing acceptor solid solubility, the Ge:Ga system appears to be favorable compared to diamond:B and Si:B [14]. Previous studies demonstrated that conventional implantation doping of Ge with Ga enables a maximum hole concentration of $6.6 \times 10^{20}$ cm$^{-3}$ after annealing at 450 °C for 1 h [24]. Higher temperatures of conventional long-term annealing led to Ga clustering. In order to reduce acceptor diffusion and clustering, flash lamp annealing (FLA) in the ms-range without layer melting is an appropriate method [25-28]. With this method hole concentrations up to $1.4 \times 10^{21}$ cm$^{-3}$ and superconductivity at critical temperatures below 0.5 K [4, 15] and 2.0 K [29] have been achieved in Ge layers with about 6 at. % and 8 at. % Ga content, respectively. Unfortunately, the layers are nanocrystalline [4, 15], and the activation level of the Ga acceptors varies from sample to sample up to a factor of two, which is due to the formation of Ga-rich nanoprecipitates [29]. Single-crystalline Ge:Ga has been obtained by RTA [14]. However, in this case, a large amount of the Ga atoms accumulates as an amorphous film at the SiO$_2$/Ge-interface. This interface layer becomes superconducting below 6 K which is similar to the critical temperature of Ga clusters.

In this paper report we show that an optimized FLA process can be used to fabricate single-crystalline, superconducting layers of hyperdoped p-type Ge. In addition to Ga doping, we also investigate Al doping. Similarly to Ga, Al has high ESS, but a higher diffusivity than Ga and is more difficult to activate. As shown recently, ion implantation of Al into Ge followed by conventional thermal annealing results in a maximum hole concentration of only $1 \times 10^{20}$ cm$^{-3}$ [30]. In this paper, it was shown that the maximum carrier concentration in Ga- and Al-implanted Ge followed by FLA exceeded $10^{21}$ cm$^{-3}$. FLA suppresses the dopant diffusion and segregation. The recrystallized Ge is single-crystalline with critical temperatures of $T_C \sim 0.5$ K. Moreover, first-principles investigation of superconductivity in Al-doped and Ga-doped Ge using ab-initio calculations within the Eliashberg/McMillan theory reveals that the Ga:Al system behaves similar to Ge:Ga covalent superconductor, where the critical temperature can be tuned by the carrier concentration.

**2. Experimental**

**2.1. Sample fabrication**

N-type (Sb-doped, $\rho > 10$ $\Omega$cm), (100)-oriented Ge wafers are used as substrates for acceptor implantation in order to electrically isolate the processed layer from the substrate by a formation of a p-n junction. First, a 30 nm thick $SiO_2$ cover layer is sputter-deposited to protect the Ge surface during ion implantation and annealing. Then, the wafers are implanted with Ga or Al ions with different fluencis of 1, 2 and $4 \times 10^{16}$ cm$^{-2}$ and energies of 100 keV for Ga$^+$ and 50 keV for Al$^+$ ions. The implantation energies are chosen in such a way that the acceptor profiles are similar with a maximum acceptor concentration at a depth of 60 nm, as predicted by the SRIM simulation code [31]. Fig. 1a shows the calculated Ga and Al distributions implanted into the $SiO_2$/Ge wafers for the ion fluence of $2 \times 10^{16}$ cm$^{-2}$. The peak concentration and the depth distribution of Al and Ga within Ge are different for the same ion fluence and similar projected ion range $R_p$. After implantation, a heavily doped amorphous surface layer of about 120 nm width with a relatively sharp interface to the single crystalline Ge substrate was formed (see Supplemental Material Fig. S1a) [15]. The presence of a sharp amorphous/crystalline interface is an important precondition for the explosive solid-phase epitaxy process which appears during ms-range FLA of the implanted layer [28].

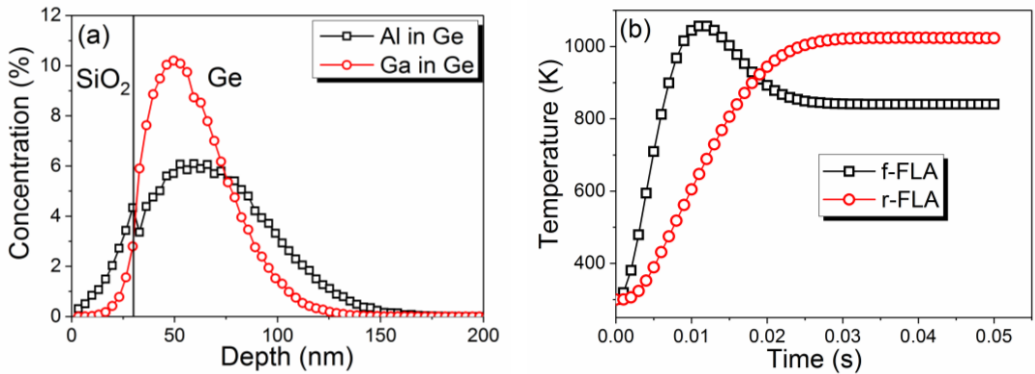

Figure 1. The Ga and Al depth distribution in Ge obtained by the SRIM code for an ion implantation fluence of $2 \times 10^{16}$ cm$^{-2}$ (a) and simulated temperature distribution at the implanted surface during 20 ms FLA from the front side (black curve) and from the rear side (red curve) (b).

The Al peak concentration is about 6 at.-%, whereas the Ga concentration exceeds 10 at.-%. This is due to different interactions of light (Al) and heavy (Ga) elements with germanium during the ion implantation process. In particular, it is due to different stopping power and energy loss straggling for different ions within the solid. For heavier ions the stopping power and the energy loss straggling are higher causing a smaller full width half maximum of the depth distribution of the implanted ions and, in consequence, a higher peak concentration for the same ion fluence.

In order to activate the dopants and recrystallize the implanted layer, we have used a strongly non-equilibrium thermal processing, i.e. flash lamp annealing. Implanted samples were annealed either from the front side (f-FLA) or from the rear side (r-FLA) with an energy density deposited to the sample surface in the range of 50 to 130 Jcm$^{-2}$. The annealing time was 3, 6 or 20 ms. The influence of the annealing time on the recrystallization process of implanted layer is presented in Supplemental Material (see Fig. S1b). Figure 1 shows the temperature distribution within implanted layer after front and rear side FLA for 20 ms. The f-FLA leads to a partial epitaxial regrowth of the implanted layer and to the formation of polycrystalline hyperdoped Ge at the surface [15]. Taking into account the wavelength spectrum of the Xe lamps in the FLA system (300 – 800 nm) and the optical properties of Ge, the main part of the flash light is absorbed by implanted Ge within 50 nm from the surface. This causes a temperature gradient within the implanted layer. For a short moment (in the sub-microsecond range), the surface is much hotter than the amorphous/crystalline interface. Also, the threshold energy needed for crystalline seed nucleation is lower than the energy needed for the epitaxial regrowth [32]. Therefore, during f-FLA, the recrystallization of the implanted layer starts from the surface and a polycrystalline layer is formed. In order to avoid the formation of such polycrystalline layer at the top of implanted Ge, we developed the rear-side FLA process [28]. In this case, the implanted sample is annealed from the rear side and the heat is transferred through the wafer to the implanted surface. Using r-FLA, the amorphous/crystalline interface is heated first. Therefore, before the surface temperature reaches the level needed for crystalline seeds nucleation, the whole implanted layer is recrystallized due to the explosive solid phase epitaxy [28]. We have found that using 400 μm thick Ge layer the optimal annealing time for rear side annealing is 20 ms. Alternative to FLA annealing techniques are rapid thermal annealing (RTA) and pulsed laser annealing (PLA). During RTA, similar to r-FLA process, the implanted layer recrystallizes via solid phase epitaxy. But due to much smaller heating rate the recrystallization speed of implanted layer is significantly slower than the diffusion of dopants. As a consequence the implanted elements with concentration higher than the solid solubility are only partially incorporated into the crystal lattice and form clusters. In contrast to FLA and RTA, during PLA the annealing layer recrystallizes via liquid phase epitaxy. The typical pulse length for PLA is in the nanosecond range and the total annealing time is in the range of tens of microseconds. The solidification/recrystallization speed observed during PLA is similar to the explosive solid phase epitaxy after FLA. But the diffusion coefficient of dopants in the liquid phase is a few orders of magnitude higher than in the solids. Hence, during PLA dopants often diffuse towards the surface and form a dopant-rich but non-activated layer. According to our experience only the millisecond range annealing provides enough energy to activate explosive solid phase epitaxy which is crucial for the formation of single crystalline hyperdoped germanium.

### 2.2. Characterization techniques

The crystallization process of the Al- and Ga-implanted and annealed samples is studied using Rutherford backscattering-channelling spectrometry (RBS/C). The RBS/C measurements are performed on the samples before and after annealing using the 1.7 MeV $He^+$ beam. To investigate the microstructural properties of the implanted Ge layer, cross-sectional bright-field transmission electron microscopy (TEM) investigations are performed in a Titan 80-300 (FEI) microscope operated at an accelerating voltage of 300 kV. High-angle annular dark-field scanning transmission electron microscopy (HAADF-STEM) imaging and spectrum imaging based on energy-dispersive X-ray spectroscopy (EDXS) are performed at 200 kV with a Talos F200X microscope equipped with a Super-X EDXS detector system (FEI). Prior to TEM analysis, the specimen mounted in a high-visibility low-background holder was placed for 10 s into a Model 1020 Plasma Cleaner (Fischione) to remove contaminations. The optical properties are investigated by micro -Raman spectroscopy. The phonon spectra were obtained in backscattering geometry in the range of 100 to 600 $cm^{-1}$ using a 532 nm Nd:YAG laser with a liquid nitrogen-cooled charge coupled device camera.

The concentration of carriers in the implanted and annealed samples was estimated from temperature-dependent Hall effect measurements in van der Pauw configuration. The thickness of the doped layer was extracted from the RBS data under the assumption that the diffusion of implanted elements during 20 ms pulse annealing can be neglected. The electrical properties of the annealed samples are measured at mK temperatures in a dry dilution refrigerator (Triton 400 by Oxford Instruments), which allows sweeping temperature in the range from 10 mK to 30 K. Four-probe AC measurements were taken using the AC lock-in method with an excitation current of 10 nA and frequency 127 Hz.

The existence of superconducting states in hyperdoped p-type Ge was predicted by ab initio calculations within the Eliashberg/McMillan theory.

## 3. Results and Discussion

### 3.1. Microstructure

The recrystallization process of ion-implanted and flash-lamp-annealed Ge is investigated using RBS random (RBS/R) and channeling (RBS/C) spectrometry. Since Al and Ga are lighter than Ge, they unfortunately cannot be measured directly by RBS. However, the ratio between the random and channeling spectra provides information about the quality of the crystal structure. The ratio between the yield of the RBS/C and RBS/R spectra ($\chi_{min}$) is a measure of the crystalline quality. In our case, $\chi_{min}$ for Al- and Ga-hyperdoped Ge after FLA is in the range of (5±1) % (see Fig. 5), which is slightly higher than $\chi_{min}$ for the virgin Ge. Moreover, the RBS/C spectra recorded from the as-implanted samples provide information about the thickness of the amorphized layer which is needed to calculate the carrier density using Hall effect measurements.

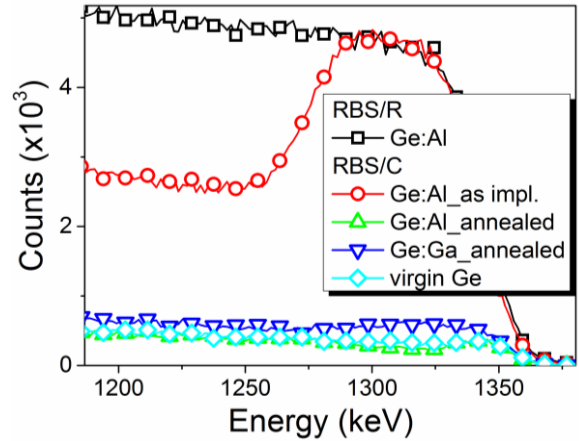

Figure 2. The RBS/R and RBS/C spectra obtained from Al- and Ga-hyperdoped Ge. The RBS/C spectrum recorded from virgin Ge is shown as well. The concentrations of Al and Ga in Ge are at the level of $2 \times 10^{21}$ cm$^{-3}$.

Figure 2 shows the RBS/R and RBS/C spectra obtained from the Al-doped sample before and after FLA, from the Ga-doped sample after FLA and from virgin Ge. In order to avoid the formation of a polycrystalline layer at the top of the sample, FLA has been performed from the rear side. As expected, the RBS/C spectrum obtained from the as-implanted sample reveals the formation of an about 120 nm thick amorphous surface layer. The thickness of the amorphous layer is calculated based on the RBS data using the RUMP Software. After 20 ms r-FLA with an energy density of 120 Jcm$^{-2}$, the yield of RBS/C spectrum drops down to the level registered from the virgin Ge wafer. This behavior points to an epitaxial regrowth of the implanted layer during FLA. Moreover, we can conclude that Al atoms are incorporated into the lattice of Ge. Taking into account that the solid solubility of Al in Ge is in the range of $5 \times 10^{20}$ cm$^{-3}$, the investigated sample contains four times more Al in substitutional positions than the solid solubility limit. Such a gain is only possible due to the strongly non-equilibrium character of the process. The absence of significant dechanneling suggests that the formation of Al clusters is also suppressed by the ms-range r-FLA.

In the case of Ga-hyperdoped Ge after r-FLA, the RBS/C spectrum also reveals full incorporation of Ga into the Ge lattice. The yield of the RBS/C spectrum obtained from virgin Ge and the Ga-doped sample is at the same level, meaning that the Ga-implanted sample behaves the same way like Al-doped Ge after r-FLA. In both cases, the ESS limit has got overcome by four times.

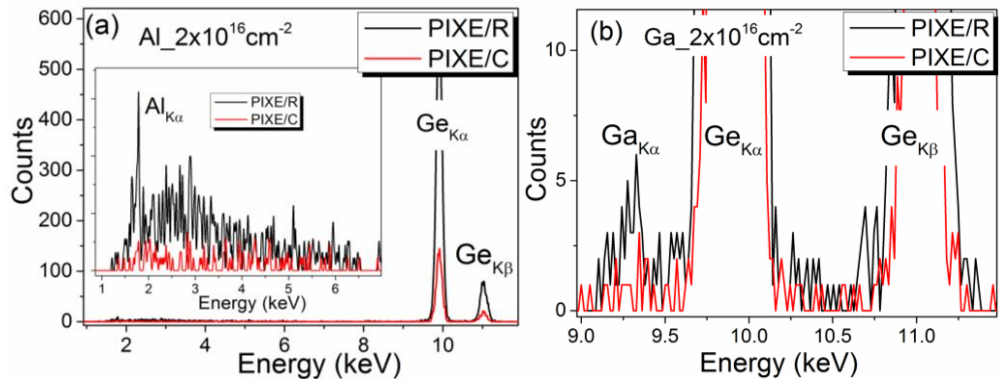

Figure 3. PIXE spectra of Al- (a) and Ga-hyperdoped Ge (b) followed by rear-side FLA for 20 ms.

In order to clarify the lattice position of Al and Ga within Ge, we performed PIXE spectroscopy in the random and channeling direction. Figure 3a and b show the PIXE spectra obtained from the Al- and Ga-doped samples, respectively. The peaks are identified as the characteristic X-ray emissions of the $Al_{K\alpha}$ (1.78 keV), $Ga_{K\alpha}$ (9.27 keV), $Ge_{K\alpha}$ (9.85 keV) and $Ge_{K\beta}$ (10.98 keV) lines. Since in the PIXE channeling spectra the $Al_{K\alpha}$ and $Ga_{K\alpha}$ intensities drop down to the noise level it can be concluded that both Al and Ga atoms are fully incorporated into Ge lattice sites even with a concentration being four times higher than the ESS.

More light on the microstructure is provided by TEM. Please note that the $SiO_2$ capping layer is still present for these samples. Figure 4a displays a cross-sectional bright-field TEM image taken from Al-doped Ge after annealing. In this case, both single dislocations within the implanted layer and end-of-range defects are detected [33]. Figure 4b shows the Ge, Al and O distributions based on EDXS analysis from a representative surface region, as exemplarily marked by the white square in Fig. 4a. Aluminum is quite evenly distributed within Ge showing only few small agglomerates over the implantation depth which is in good agreement with RBS and PIXE data.

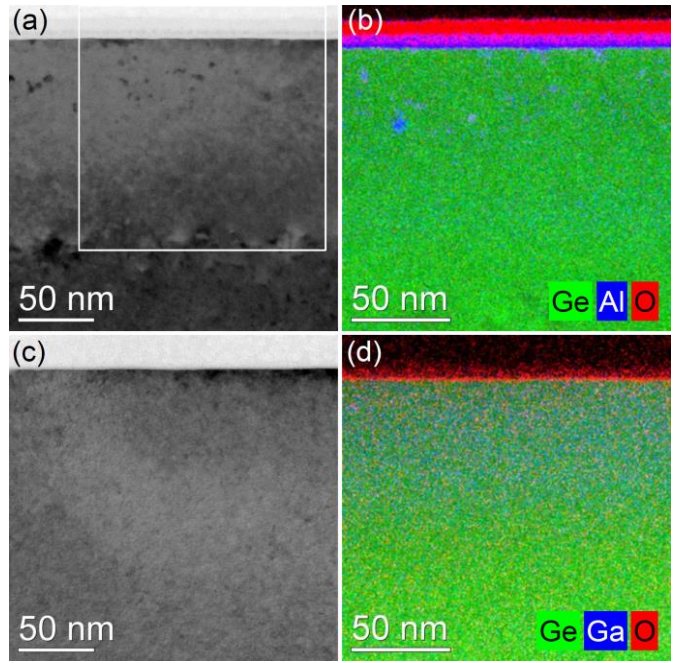

Figure 4. (a) and (c) Cross-sectional bright-field TEM images obtained from Al- and Ga-hyperdoped Ge, respectively, (b) and (d) superimposed Ge (green), O (red) and Al or Ga (blue), respectively, element distributions obtained by spectrum imaging analysis based on EDXS in scanning TEM mode for a representative surface region of each sample, as exemplarily marked by the white square in (a).

Figures 4c and d show a cross-sectional bright-field TEM micrograph and the corresponding superimposed Ge, Ga and O element distributions obtained from Ga-hyperdoped Ge. Here, the Ga is completely homogeneously distributed within the implanted layer. Moreover, in the case of the Ga-doped sample, even the end-of-range defects are not detected. For Al as well as Ga, the recrystallized Ge is single-crystalline. This is in contrast to our previous results where front-side flash lamp annealing was used [15]. Applying f-FLA, the implanted layer is composed of polycrystalline Ge with Ga clusters and an epitaxial layer which has a thickness of about 70 % of the thickness of the implanted layer. Using r-FLA, we can fully suppress the formation of poly-Ge and Ga clusters for dopant concnetrations much above the ESS.

### 3.2. Superconductivity

The established electrical parameters of the studied doped layers are summarized in Table 1. The carrier concentration was estimated from the Hall Effect measurements. The thickness of the doped layer was determined by RBS measurements. The presented activation efficiency is a ration between the total acceptor concentration and the carrier concertation estimated from the Hall effect measurement at 3 K. The presented critical temperatures are taken from Fig. 5.

We find that for diluted hyperdoped Ge the minimum carrier concentration needed for superconductivity is in the range of $1\times10^{21}$ cm$^{-3}$. If we assume that the superconductivity in p-type Ge is phonon-mediated, for the same doping level the critical temperature should be slightly higher for Al-doped Ge than for Ga-doped sample due to a stronger phonon coupling. Unfortunately, in hyperdoped Ge it is very challenging to control the carrier concentration keeping the same dopant concentration. In fact, for the doping level above the solid solubility, we are not able to activate 100% of the implanted element. Therefore experimental verification of theoretical predictions is very challenging. Here we decided to compare samples with similar hole concentration.

Tabele 1. Summary of the doping level and carrier concentration in hyperdoped Ge obtained for Al and Ga doped samples annealed with optimized parameters.

| Sample | Dopant concentration | Carrier concentration at 3K | Activation efficiency | Critical temp. |
|---|---|---|---|---|
| Ge:Al | ~6% | $10.7\times10^{20}$ cm$^{-3}$ | 44.5% | $T_C$~0.15K |
| Ge:Ga | ~10% | $12.6\times10^{20}$ cm$^{-3}$ | 31.5 % | $T_C$~ 0.45K |

It is worthy to note that, we are able to show for the first time superconducting Ge hyperdoped with Al. The achieved hole concentration is the highest ever published for Al-doped samples. Figure 5a shows the temperature dependence of resistance for Al-hyperdoped Ge. The superconducting temperature is about 150 mK. This is much lower than that predicted by calculation ($T_C$ ~480 mK shown later), but also much below the critical superconducting temperature for Al thin film or Al clusters [34]. If we take into account the carrier concentration which is roughly half of the Al concentration, the obtained $T_C$ is at a reasonable level. The temperature dependence of the resistance for the Ga-doped sample is shown in Fig. 5b. In Ga-hyperdoped Ge, $T_C$ is about 400 mK. Presented critical temperatures are obtained from samples annealed from the rear-side with a flash energy density of 120 Jcm$^{-2}$. The annealing at lower energy densities is not sufficient to recrystallize the implanted layer (no superconductivity), while annealing at higher energy densities activates the dopant diffusion and cluster formation leading to the superconductivity driven by metallic clusters (see Supplemental Material Figs. 1b and 2) [15]. In consequence, the fraction of electrically active dopants, both Al and Ga in the substitutional position, is much smaller.

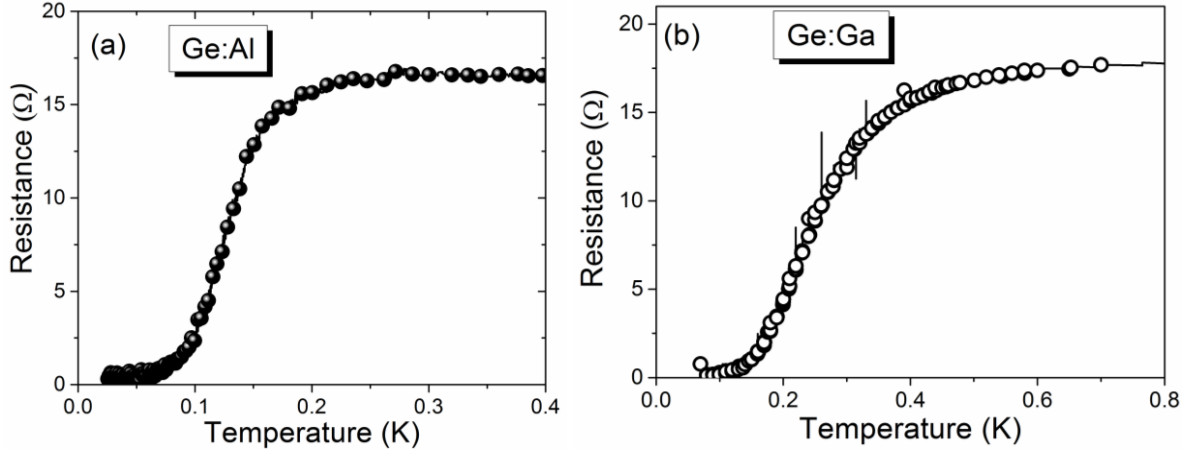

Figure 5. Temperature dependence of the longitudinal resistance for different samples: (a) low temperature part of $R_{xx}$ vs T of superconducting Ge with 6 % of Al (a) and Ga (b).

## 3.3. Model calculations for the electron-phonon coupling

According to the BCS theory the critical temperature of a homogeneous superconductor grows with increasing electron-phonon coupling strength and Debye temperature. Theoretical calculations demonstrate that in homogeneously doped semiconductors the critical temperature scales with their hole concentration [35, 36]. The critical temperature in diamond can exceed 20 K for a hole concentration of 10% (~$10^{22}$ cm$^{-3}$). Much lower critical temperatures (< 1 K) have been predicted for Si and Ge. We used the supercell technique to model the hyperdoped Ge. For simulation we have used Ge doped with Al or Ga with the concentration of 6.25 %, corresponding to the 2×2×2 supercell with one Ge atom substituted by an Al(Ga) atom. All calculations were performed within the plane-wave implementation of the local density approximation (LDA) [37] to density-functional theory (DFT) [37-39] in the QUANTUM-ESPRESSO package [40]. Norm-Conserving pseudopotentials with a kinetic energy cutoff of 45 Ry were used to represent electron-ion interactions. The k-point sampling of the Brillouin zone was set to 6×6×6 during the structural relaxation and electronic structure calculations, while a dense 12×12×12 Monkhors-Pack grid [31] was used for the phonon linewidth calculations. Phonon spectra and electron-phonon coupling constants were calculated using density-functional perturbation theory [41] with an 3×3×1 mesh of q points. For all calculation we have used an optimized lattice constant of Ge supercell of 11.234 Å. The hyperdoping of Ge with Al or Ga will lead to a lattice expansion by 0.2 % (11.259Å for Al doped Ge) or lattice compression by 0.1 % ( 11.226 Å for Ga doped Ge) , respectively.

Figure 6 shows the electronic structure of Al and Ga hyperdoped Ge. According to our calculations for the same dopant concentration the density of states $N(E_F)$ in Al hyperdoped Ge (~2.82[states/eV/(supercell)]) is slightly higher than in Ga hyperdoped Ge (~2.67[states/eV/(supercell)]). In both cases the hyperdoped Ge is strongly degenerate with the Fermi level ($E_F$) located deep in the valence band (~ 0.67 eV below the top of valence band maximum). The electronic states near the $E_F$ of Al and Ga hyperdoped Ge are very similar and they originate from p states of Ge and acceptor dopants. The hyperdoped Ge is an sp3 covalent metals.

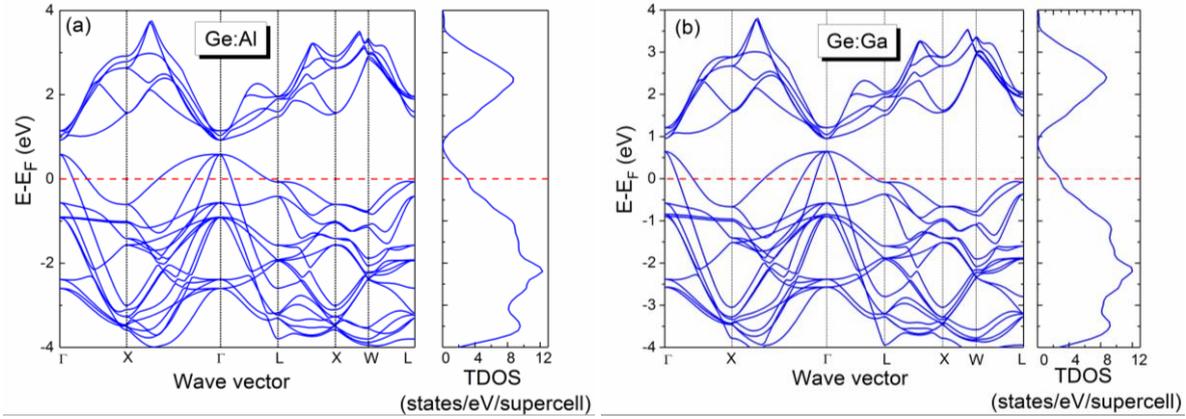

Figure 6. The electronic structure of Al (a) and Ga (b) hyperdoped Ge with the corresponding total density of electronic states $N(E_F)$.

In a similar way to the electronic structure of hyperdoped Ge we have calculated the phonon band structure of Al and Ga hyperdoped Ge (see Fig. 7). By solid blue circles we have marked the highest optical phonons at the Γ point. In the case of Al hyperdoped Ge we have found three optical $T_2$ modes located at 256 cm$^{-1}$, 267 cm$^{-1}$, 364 cm$^{-1}$ and one $A_1$ optical mode with frequency around 188 cm$^{-1}$. The strongest electron-phonon coupling strength $\lambda_{vq}$ at the Γ point is at $A_1$ with the $\lambda_{vq}$ ~ 0.16. The triple degeneracy also produces electron-phonon coupling strength of about $\lambda_{vq(q=\Gamma)}$~0.05-0.06. The phonon structure of Ga hyperdoped Ge is very similar to Al hyperdoped Ge. The $A_1$ optical mode should be located at 187 cm$^{-1}$ with maximum electron-phonon coupling of about 0.12. The optical $T_2$ modes in Ga hyperdoped Ge are located at 254 cm$^{-1}$, 259 cm$^{-1}$ and 266 cm$^{-1}$ and the $\lambda_{vq(q=\Gamma)}$ is in the range of 0.04-0.07. The theoretical calculated phonon structure of hyperdoped Ge was verified using micro-Raman spectroscopy. The micro-Raman spectra were collected under 532 nm laser excitation with the laser power of 3.2 mW and the focal diameter of about 1 μm. Figure 7c and d show the Raman spectra obtained from Al and Ga hyperdoped Ge, respectively. The transverse optical phonon mode of Ge-Ge in intrinsic Ge is located at 300.5 cm$^{-1}$. After hyperdoping the TO phonon mode of Ge-Ge vibrational is shifted down to 288.2 cm$^{-1}$ for the Al doped sample and down to 281.9 cm$^{-1}$ for the Ga doped Ge. This is very close to the theoretically predicted values for the high frequency of the zone-center optical mode in hyperdoped Ge (about 278 cm$^{-1}$). The shift of the TO phonon mode in ultra-high doped Ge and the peak asymmetry is due to the phonon softening and the Fano effect [42-44]. Beside the TO phonon mode we can easily distinguish the $A_1$ phonon mode in both samples. The measured peak position of the $A_1$ mode located at about 188 cm$^{-1}$ fits well to the theoretically predicted phonon energy using DOS calculation.

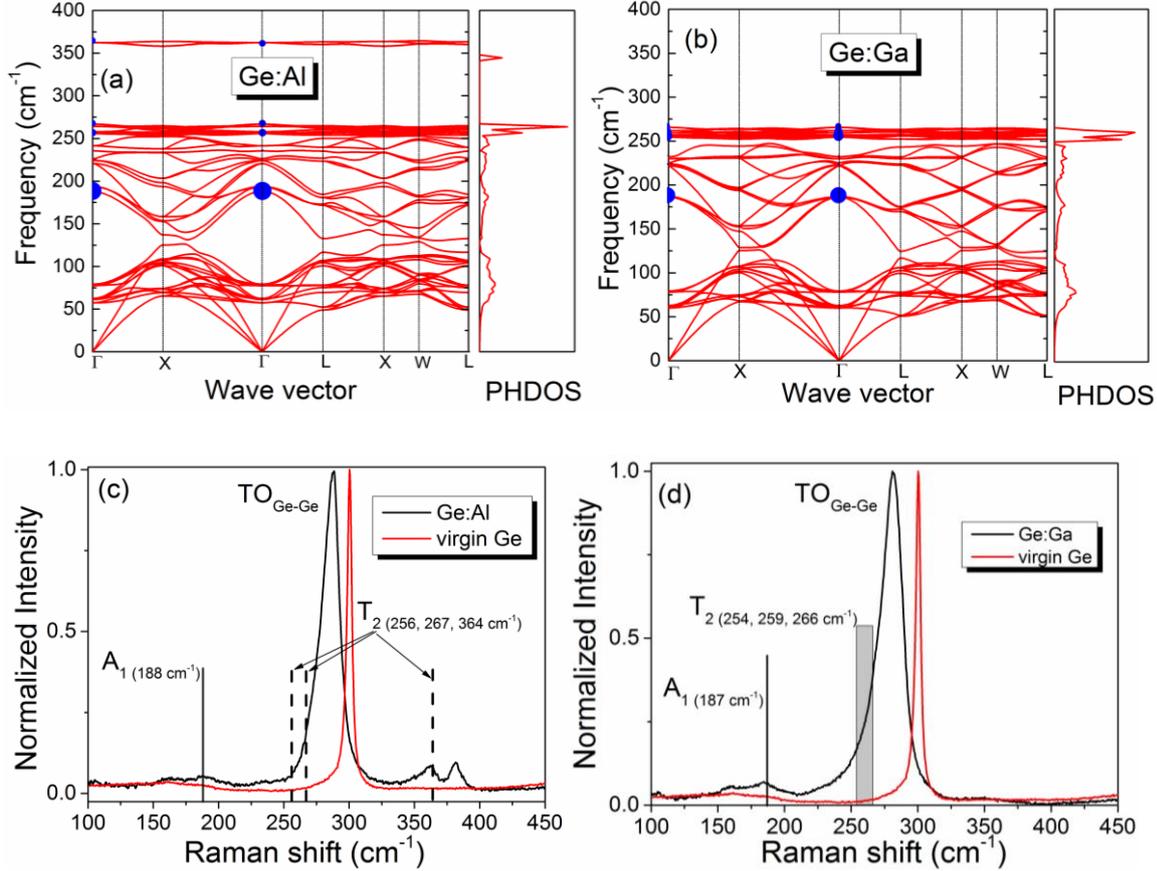

Figure 7. Phonon band structure with corresponding phonon DOS for Al hyperdoped Ge (a) and for Ga hyperdoped Ge (b). The unit of phonon DOS (PHDOS) is states/(cm$^{-1}$×supercell). Blue solid circles indicate the partial electron-phonon coupling strength $\lambda_{vq}$ at the Γ point. (c) and (d) show the Raman spectra obtained from Al and Ga hyperdoped Ge, respectively. The Raman spectrum of virgin undoped Ge is shown for comparison.

Due to the fact that the $T_2$ phonon mode positions are close to the strongest zone-center TO phonon mode it is difficult to distinguish them. But in the case of Al hyperdoped Ge the $T_2$ mode at 364 cm$^{-1}$ is well visible (see Fig. 7c). In Ga hyperdoped Ge all three $T_2$ modes are overlapped with the TO phonon mode. Next we analyze the electron-phonon coupling. Figure 8 a and b show the Eliashberg spectral function and the integrated electron-phonon coupling constant $\lambda(\omega)$. The total $\lambda$ calculated for Al and Ga doped Ge are similar and equal to 0.355 and 0.350, respectively. The calculated logarithmic phonon frequency $\omega_{log}$ is about 243.6 K for Al doped Ge and about 245.1 K for Ga doped Ge which is much smaller than the $\omega_{log}$ in other group IV superconductors like; around 700 K for Si:B and about 1287 K for boron doped diamond [44]. Finally, we have calculated the superconducting critical temperature for both samples. We found that the expected $T_C$ for Al doped Ge should be slightly higher than that for Ga doped sample, mainly due to slightly higher phonon-carrier coupling. The $T_C$ for Ge:Al is 0.48 K and for the Ge:Ga system the $T_C$ should be about 0.43 K. According to our calculation(s) the superconductivity in diluted p-type hyperdoped Ge should be phonon- mediated.

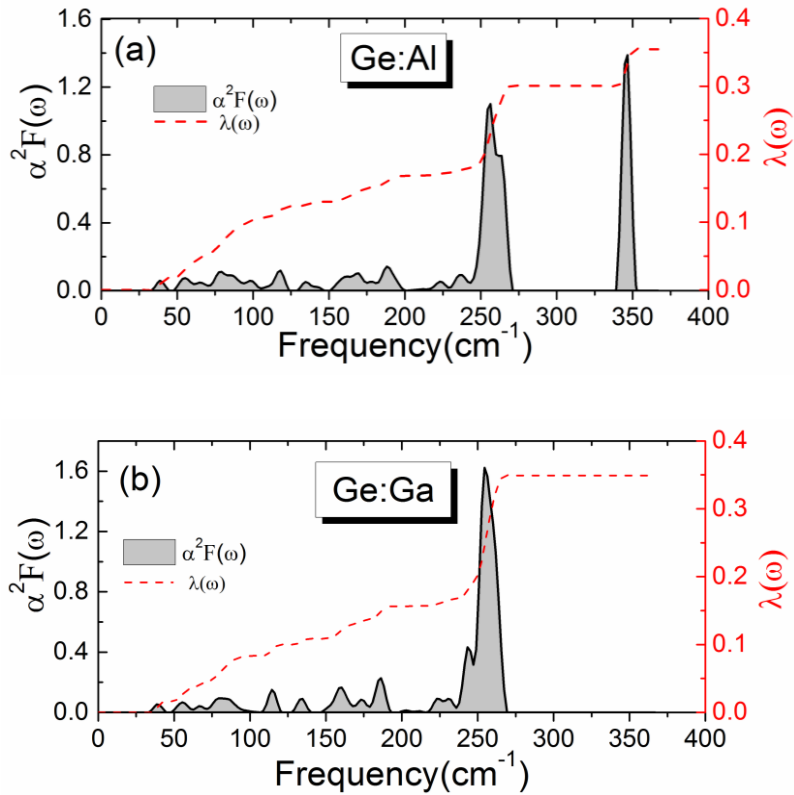

Figure 8. The total and projected Eliashberg spectral function ($\alpha^2F(\omega)$) for Al hyperdoped Ge (a) and for Ga hyperdoped Ge (b). The red dashed curves represent the integrated electron-phonon coupling constant $\lambda\ (\omega)$.

Note that there is significant discrepancy between calculated $T_C$ and the experimental values shown in Table I. The theoretical calculation cannot take into account all phenomena which may exist in real samples. In fact, we have a Gaussian distribution of the implanted elements introducing a kind of inhomogeneity into the doped layer which is not accounted in our calculations/modeling. Next, although the thermal treatment is very short, and we were not able to detect big metallic clusters within the implanted Ge layer, we cannot exclude the formation of Ga clusters with a diameter below the resolution limit of our TEM system. Moreover, the effective carrier concentration is lower than the nominal dopant concentration which is not taken into account for simulation.

**Conclusions**

We have fabricated single crystalline Al and Ga doped superconducting Ge where the diffusion and clustering of dopants are suppressed by utilization of the strongly non-equilibrium thermal processing. Using rear-side FLA the implanted Ge layers recrystallize epitaxially due to the explosive solid-phase epitaxy. The theoretically predicted critical temperatures qualitatively agree with experimental values. With further optimizing dopant concentrations and annealing parameters, our work will pave the monolithic integration of superconducting nanocircuits in semiconductor devices.

**Acknowledgement**

Support by the Ion Beam Center (IBC) at HZDR and the funding of TEM Talos by the German Federal Ministry of Education of Research (BMBF), Grant No. 03SF0451 in the framework of HEMCP are gratefully acknowledged. We would like to thank Andrea Scholz for XRD measurements and Romy Aniol for TEM specimen preparation. This work was partially supported by the German Academic Exchange Service (DAAD, Project-ID:57216326) and National Science Centre, Poland, under Grant No. 2016/23/B/ST7/03451. The work was partially supported by the EU 7th Framework Programme project REGPOT-CT-2013-316014 (EAgLE) and by the Foundation for Polish Science through the IRA Programme co-financed by EU within SG OP.

# Supplemental Material

# Superconductivity in single-crystalline, aluminum- and gallium-hyperdoped germanium

Slawomir Prucnal[1], Viton Heera[1], René Hübner[1], Mao Wang[1], Grzegorz P. Mazur[2,3], Michał J. Grzybowski[2,3], Xin Qin[4], Ye Yuan[1,5], Matthias Voelskow[1], Wolfgang Skorupa[1], Lars Rebohle[1], Manfred Helm[1], Maciej Sawicki[3], Shengqiang Zhou[1]

*1. Helmholtz-Zentrum Dresden-Rossendorf, Institute of Ion Beam Physics and Materials Research, Bautzner Landstraße 400, D-01328 Dresden, Germany*

*2. Institute of Physics, Polish Academy of Sciences, Aleja Lotnikow 32/46, PL-02668 Warsaw, Poland*

*3. International Research Centre MagTop, Institute of Physics, Polish Academy of Sciences, Aleja Lotnikow 32/46, PL-02668 Warsaw, Poland*

*4. Hefei National Laboratory for Physical Sciences at the Microscale, University of Science and Technology of China, Hefei, Anhui, 230026, China.*

*5. Physical Science and Engineering Division, King Abdullah University of Science and Technology, 23955-6900 Thuwal, Saudi Arabia*

## 2. Experimental

### 2.1. Sample fabrication

N-type (Sb-doped, $\rho > 10$ $\Omega$cm), (100)-oriented Ge wafers are used as substrates for acceptor implantation in order to electrically isolate the processed layer from the substrate by the formation of a p-n junction. First, a 30 nm thick $SiO_2$ cover layer is sputter-deposited to protect the Ge surface during ion implantation and annealing. Then samples were implanted with Al and Ga ions. Al was implanted with two different fluences of 2 and $4\times10^{16}$ cm$^2$ and an energy of 50 keV. Ga ions were implanted with three fluences of 1, 2 and $4\times10^{16}$ cm$^2$ and an energy of 100 keV. After ion implantation, samples were treated by flash lamp annealing (FLA) using different annealing times (3, 6 and 20 ms) and energy densities deposited onto the sample surface (from 50 up to 130 Jcm$^{-2}$). We have also tested the front- and rear-side annealing.

According to our previous experiments for Ga-implanted Ge [1], front-side FLA leads to the formation of a polycrystalline layer. The same effect was observed for Al-implanted samples. After front FLA, the implanted layer consists of an about 30 nm thick polycrystalline layer and a 70 nm thick single-crystalline layer which is formed due to explosive solid phase epitaxy [2]. The influence of FLA on the recrystallization process of ion-implanted Ge is explained with details in Ref. 2. Therefore, we focused here on rear-side annealing only. In order to recrystallize the implanted layer, the energy deposited into the rear side must be transferred through the entire wafer to the front side. The thermal conductivity of Ge determines the minimum annealing time needed to recrystallize the implanted layer. The maximum energy density deposited onto the sample surface (peak temperature) is limited by the melting temperature. For a 400 μm thick Ge wafer, the optimum annealing time was found to be 20 ms, both for Al- and for Ga-implanted samples.

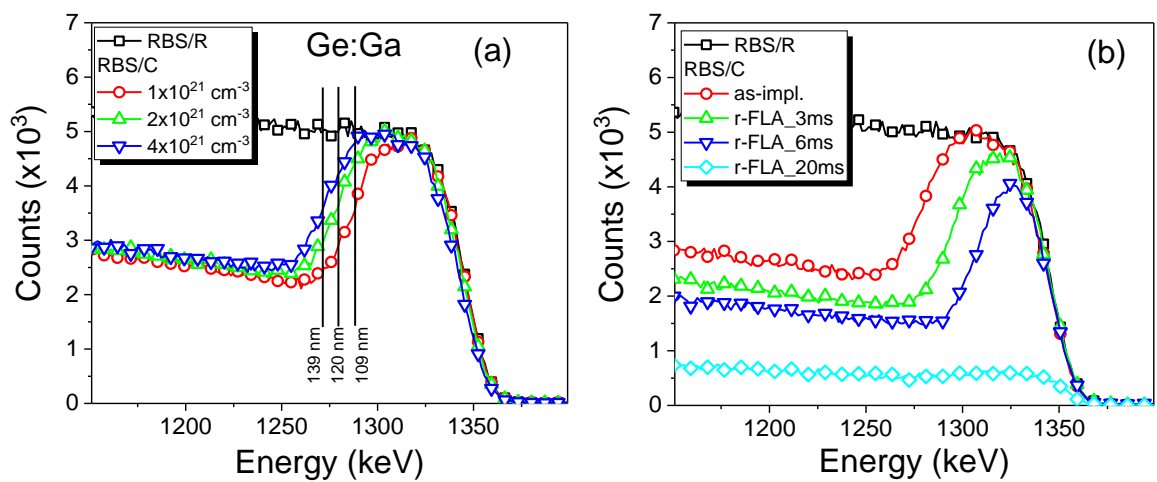

Figure S1. The RBS/R and RBS/C spectra obtained from Ga-hyperdoped Ge. (a) shows the RBS spectra obtained from the as-implanted samples with Ga concentrations at a level of 1, 2 and $4 \times 10^{21}$ cm$^{-3}$, whereas the amorphous layer thicknesses are given for each case. (b) shows the influence of the pulse duration on the recrystallization of Ga-implanted Ge with a Ga concentration of $2 \times 10^{21}$ cm$^{-3}$.

Figure S1a shows random (RBS/R) and channelling Rutherford backscattering (RBS/C) spectra obtained from as-implanted Ge samples containing three different Ga concentrations. The thicknesses of the amorphous layers for each Ga concentration is calculated based on the RBS data using the RUMP Software [3]. Later, these thicknesses were used to calculate the effective carrier concentration measured using the Hall Effect. The broadening of the amorphous layer with increasing Ga concentration is due to the high-fluence implantation process (elastic scattering between already implanted and new ions). In the case of Al, we have used only two concentrations of 2 and $4 \times 10^{21}$ cm$^{-3}$. The RBS data are similar to those presented in Fig. S1a. The Al-implanted Ge samples were annealed with conditions optimized for Ga-doped Ge, therefore, we expect a comparable evolution of the recrystallization process during FLA. Figure S1b shows the evolution of the explosive solid phase epitaxy as a function of annealing time. The melting point of Ge determines for each pulse length the maximum energy density which can be deposited onto the rear side of the implanted sample. For each annealing time, the flashed side was fully or partially molten. This means that the peak temperature at the flashed surface (rear side) reaches 938 °C. When the annealed surface becomes liquid, its optical reflectivity increases dramatically due to the strong increase of the amount of free equilibrium carriers in a molten crystal [4]. Therefore, using rear-side flash lamp or laser annealing, the temperature at the front side depends on the flash/laser parameters, thickness and thermal conductivity of the annealed material. As can be seen in Figure S1b, after FLA for 3 and 6 ms, only a part of the implanted Ge recrystallizes. The full recrystallization of the implanted layer appears after FLA for 20 ms. After annealing for 20 ms, the temperature at the implanted surface is in the range of 850 °C (see Fig. 1b in the main text). Importantly, the total heating/cooling process takes less than 100 ms. Using 3 ms pulse length, the temperature difference between the rear and front sides is above 250 K, while the explosive solid phase epitaxy takes place when the temperature of the annealed material is close to the melting point. Therefore, after 3 or 6 ms pulse length the implanted Ge layer started to recrystallize via conventional solid phase epitaxy which is to slow to recrystallize 120 nm within few milliseconds. In principle, the explosive solid phase epitaxy is possible also for short flash pulses (below 6 ms). In order to reduce the temperature gradient between the rear side and the front side, thinner wafers are needed.

The microstructural properties of the implanted and annealed samples were investigated by transmission electron microscopy (TEM). Figure S2 shows cross-sectional bright-field images obtained from Al- and Ga-hyperdoped Ge with the highest dopant concentration ($4 \times 10^{21}$ cm$^{-3}$) after FLA as well as the corresponding element distributions obtained by spectrum imaging analysis based on energy-dispersive X-ray spectroscopy (EDXS) in scanning TEM mode for a representative surface region of each sample, as marked by the gray squares in Fig. S2a and c. In both samples, the bright-field images reveal the presence of amorphous inclusions within the recrystallized Ge (Figs. S2a and c). High-resolution TEM images (not shown here) confirm this observation. In the case of the Al-implanted sample, the Al distribution within Ge agrees well with the spherically shaped amorphous inclusions. Additionally, a well-defined aluminum oxide layer is found at the Ge/SiO$_2$ interface (Fig. S2d). For the Ga-implanted sample, such interface layer is not observed. Rather, most of the Ga is evenly distributed within the recrystallized Ge, while an additional fraction is segregated in spherical clusters (Fig. S2b). However, these Ga-rich clusters do not necessarily coincide with the irregularly shaped amorphous inclusions. .

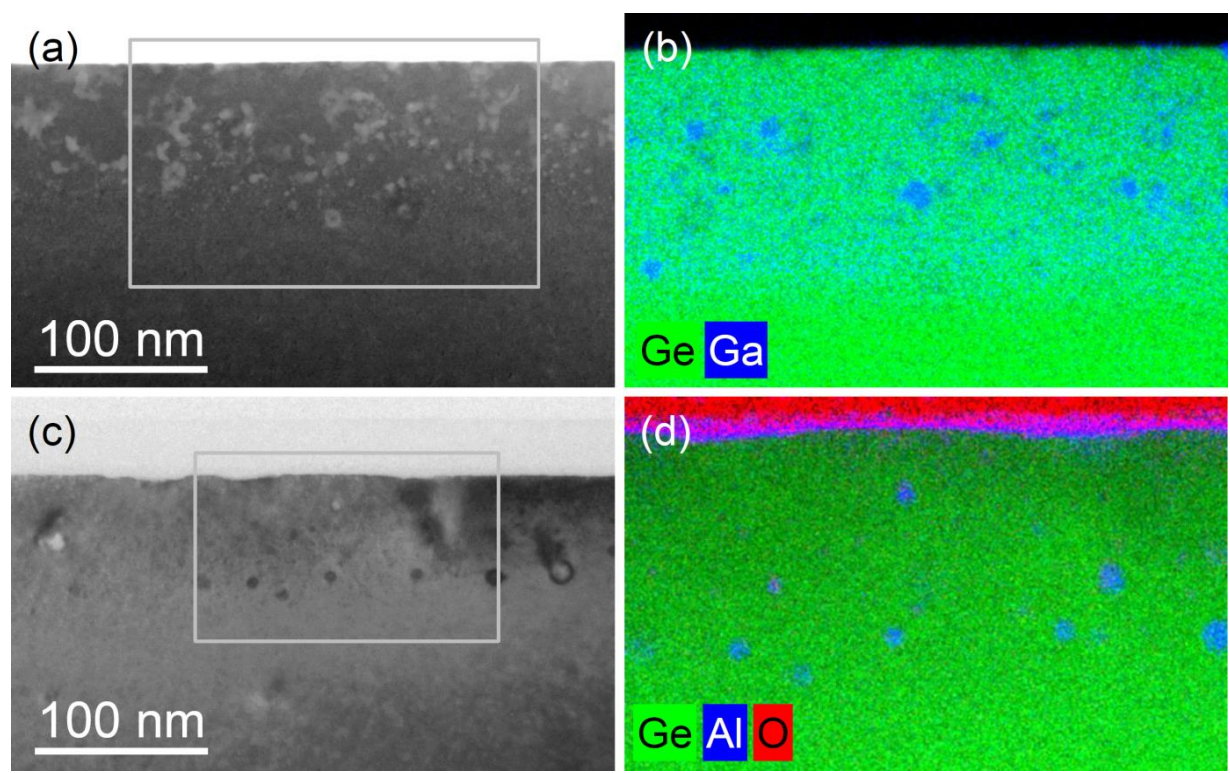

Figure S2. (a) and (c) Cross-sectional bright-field images obtained from Ga- and Al-hyperdoped Ge after FLA, respectively, (b) and (d) superimposed Ge (green), O (red) and Al or Ga (blue), respectively, element distributions obtained by spectrum imaging analysis based on EDXS in scanning TEM mode for a representative surface region of each sample, as marked by the squares in (a) and (c).

The aluminum oxide and SiO$_2$ capping layers were removed before electrical contact formation, which is why these oxides do not affect the electrical measurements. In summary, according to RBS and cross-sectional TEM investigation, we have determined that the maximum acceptor concentration which can be incorporated into the Ge crystal without cluster formation is in the range of $2 \times 10^{21}$ cm$^{-3}$, both for Al and for Ga. Figure S3 shows the sheet resistance as a function of temperature for Al- and Ga-hyperdoped Ge with different dopant concentrations. For the lowest doping level, the investigated samples show a decrease of resistance down to 80 K, followed by a slight increase up to 30 K and a stabilization for temperatures below 30 K. In the case of Ge samples containing 6 at.% of Al or more than 10 at.% of Ga, the sheet resistance continuously decreases with decreasing temperature.

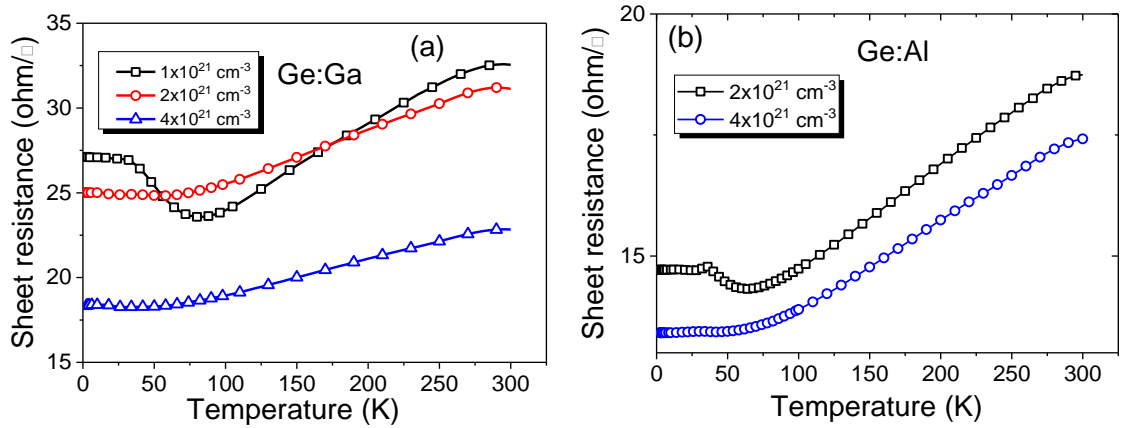

Figure S3. Sheet resistance as a function of temperature in the temperature range of 3 to 300 K for Ga (a) and Al (b) doped Ge after FLA for 20 ms with an energy density of 120 Jcm$^{-2}$.

The continuous decrease of sheet resistance with decreasing temperature is typical for metal-like samples. All samples with dopant concentration equal to or higher than $2 \times 10^{21}$ cm$^{-3}$ show superconductivity. The critical temperature ($T_C$) increases above 1 K with increasing acceptor concentration, but the effective carrier concentration doesn't increase significantly. Moreover, cross-sectional TEM reveals cluster formation for the samples with the highest dopant concentration. Hence, we concluded that the Ga-rich and Al-rich nanoprecipitates are responsible for the high-temperature superconductivity ($T_C$>1 K) in Ge.